\documentclass{INTERSPEECH2023}

\interspeechcameraready

\usepackage{cite}
\usepackage{amsmath,amssymb,amsfonts}
\usepackage{algorithmic}
\usepackage{graphicx}
\usepackage{textcomp}
\usepackage{multirow}
\usepackage[table,xcdraw]{xcolor}
\usepackage{placeins}
\usepackage{tikz}
\usetikzlibrary{spy,calc}
\usepackage{verbatim}
\usepackage{booktabs,array} 
\usepackage{url}

\usepackage[ruled,vlined]{algorithm2e}
\usepackage{nicefrac}
\usepackage{pgfplots}
\pgfplotsset{compat=newest, legend style={at={(1,0.05)},anchor=south east}
}

\usepackage{hyperref}

\hypersetup{hidelinks,
backref=true,
pagebackref=true,
hyperindex=true,
breaklinks=true,
urlcolor=blue,
bookmarks=true}

\usepackage{cleveref}

\usepackage{subfig} 

\crefname{section}{Sec.}{Sections}
\crefname{figure}{Fig.}{Figure}
\crefname{table}{Table}{Table}
\crefname{equation}{Eq.}{Equation}

\title{Federated learning for secure development of AI models for Parkinson’s disease detection using speech from different languages
\thanks{$^*$ STA and CDRU contributed equally to this work \\
Accepted for INTERSPEECH 2023, Dublin, Ireland}}

\name{Soroosh Tayebi Arasteh$^{*,1,2,3}$, Cristian David Rios-Urrego$^{*,4}$, Elmar Noeth$^{1}$, Andreas Maier$^{1}$, Seung Hee Yang$^{2}$, Jan Rusz$^{5}$, Juan Rafael Orozco-Arroyave$^{1,4}$}

\address{
  $^1$Pattern Recognition Lab, Friedrich-Alexander-Universität Erlangen-Nürnberg, Erlangen, Germany\\
  $^2$Speech \& Language Processing Lab, Friedrich-Alexander-Universität Erlangen-Nürnberg, Erlangen, Germany\\
    $^3$Department of Diagnostic and Interventional Radiology, University Hospital RWTH Aachen, Aachen, Germany\\
    $^4$GITA Lab, Faculty of Engineering, University of Antioquia, Medellín, Colombia\\
    $^5$Department of Circuit Theory, Czech Technical University in Prague, Prague, Czech Republic}
    \email{soroosh.arasteh@fau.de, cdavid.rios@udea.edu.co}

\begin{document}

\maketitle
%

\begin{abstract}

Parkinson's disease (PD) is a neurological disorder impacting a person's speech. Among automatic PD assessment methods, deep learning models have gained particular interest. Recently, the community has explored cross-pathology and cross-language models which can improve diagnostic accuracy even further. However, strict patient data privacy regulations largely prevent institutions from sharing patient speech data with each other. In this paper, we employ federated learning (FL) for PD detection using speech signals from 3 real-world language corpora of German, Spanish, and Czech, each from a separate institution. Our results indicate that the FL model outperforms all the local models in terms of diagnostic accuracy, while not performing very differently from the model based on centrally combined training sets, with the advantage of not requiring any data sharing among collaborators. This will simplify inter-institutional collaborations, resulting in enhancement of patient outcomes.

\end{abstract}
\noindent\textbf{Index Terms}: federated learning, speech pathology, Parkinson's disease, deep learning, trustworthy speech processing

\section{Introduction}
\label{sec:introduction}

Parkinson's disease (PD) is a neurodegenerative disorder that affects the nervous system, leading to the progressive deterioration of motor and non-motor functions, which contribute significantly to decreasing the quality of life of the patient's~\cite{logemann1978frequency}. PD is characterized by resting tremor, rigidity, bradykinesia, postural instability, and other symptoms~\cite{mckinlay2008profile}. Most PD patients develop speech deficits which are grouped and called hypokinetic dysarthria where the speech is characterized by reduced loudness, monotonous pitch, and changes in voice quality~\cite{pinto2004treatments,spencer2005speech}. Speech signals can be analyzed objectively to quantify the severity of the disease and track its progression over time, which can be useful in clinical research and treatment monitoring. Among the best motivations to consider the speech signals is that they can be easily collected and analyzed remotely, which can provide greater convenience to patients and reduce the need for frequent clinical visits~\cite{robin2020evaluation}. This can be especially beneficial for patients who live in remote areas or have limited mobility. In addition, speech signals can provide a complementary source of information to clinical assessment and other diagnostic tests, which can improve the accuracy and reliability of PD diagnosis and treatment~\cite{moro2021advances}.

Recently, deep learning (DL)-based methods have particularly gained a lot of attention for analyzing PD speech signals \cite{wav2vac1, wav2vac2}. However, a major impediment to developing such robust DL models is the need for accessing lots of training data, which is challenging for many institutions. Thus, benefiting from data from different external institutions could solve this issue. However, strict patient data privacy regulations in the medical context make this infeasible in most cases in real-world practice \cite{nautsch2019preserving, kaissis2021end, kaissis2020secure, privatefiarsta}. Therefore, privacy-preserving collaborative training methods, in which participating institutions do not share data with each other are favorable. Federated learning (FL) \cite{Konecn2016FederatedOD, konevcny2016federated, mcmahan2017communication}, as the golden key to this issue, has been increasingly investigated by researchers and practitioners and received a lot of attention in the medical image analysis domain \cite{Truhn2022.07.28.22277288, kaissis2020secure, FFLcollabst, sheller2020federated} as it does not require sharing any training data between participating institutions in the joint training process. To the best of our knowledge, collaborative training methods based on FL have not been addressed in the literature on pathological speech signals yet, despite the availability of similar privacy regulations and restrictions as in the imaging domain~\cite{nautsch2019preserving}, especially considering recent literature revealing the vulnerability of pathological speech signals in terms of patient data \cite{speechpathasv, orozco2021there, TOMASHENKO2022101362}.

In this paper, for the first time, we investigate the applicability of FL in the privacy-preserving development of DL methods for PD detection using speech signals from three real-world language corpora, each from a separate and independent institution.
We hypothesize that utilizing FL will substantially increase the diagnostic performances of networks for each local database while preserving patient privacy by avoiding data sharing between the institutions. Moreover, we assume that the FL model will perform relatively similarly, with only slight degradation compared to the hypothetical and non-privacy-preserving scenario where all the institutions could combine their training sets at a central location.

\section{Material and Methods}
\label{sec:methods}

\subsection{Methodology} 

The methodology addressed in this study consists of the following main stages: data were acquired in different languages (German, Spanish, and Czech), after, embeddings were extracted from speech signals for each participant using a pre-trained Wav2Vec 2.0 model \cite{wav2vac1, wav2vac2}, then the extracted embeddings were utilized for the secure FL training of a classification architecture, and finally, a copy of the global model is sent back to each participating site for the classification of PD patients from healthy control (HC) subjects. This methodology is summarized in \cref{fig:methodology}. Details of each stage are presented below.

\begin{figure*}
    \centering
    \includegraphics[width=\linewidth, trim={0cm, 0cm, 0cm, 0cm}, clip]{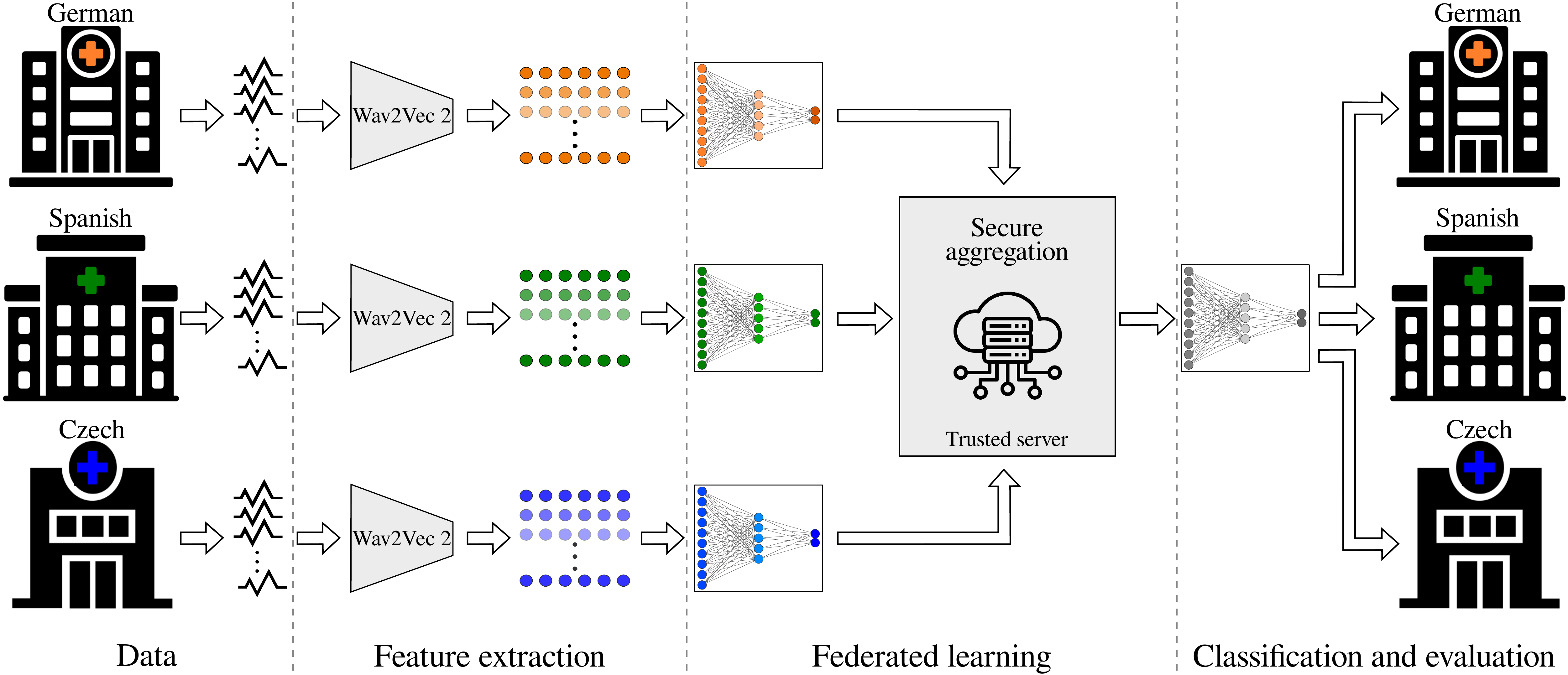}    \caption{General methodology: each institution pre-processes its local data, extracts the features using a Wav2Vec 2.0 model, and performs one epoch of the classifier network training locally, and transmits its local network parameters to a trusted server. The server aggregates all the parameters from all the institutions and transmits back the resulting global model to each institution for the next round of local training. In the end, each institution takes a copy of the final global model and performs its desired classification locally.}
    \label{fig:methodology}
\end{figure*}

\subsubsection{Data} 

We considered speech corpora in three different languages: Spanish, German, and Czech; each database contains PD patients and HC subjects. The first corpus is PC-GITA which includes recordings of 50 PD patients and 50 HC subjects~\cite{orozco2014new}. All participants were Colombian native speakers. The second corpus contained a total of 176 German native speakers (88 PD patients and 88 HC subjects)~\cite{bocklet2013automatic}. The last database contained recordings of 100 Czech native speakers divided into 50 PD patients and 50 HC subjects~\cite{rusz2018detecting}. 
Specialized neurologists evaluated each patient according to the Movement Disorder Society - Unified Parkinson's Disease Rating Scale
(MDS-UPDRS-III)~\cite{goetz2008movement}. In addition, all recordings were captured in noise-controlled conditions, and the speech signals were down-sampled to 16\,kHz to feed a deep-learning model. The rapid repetition of the syllables /pa-ta-ka/ was considered in this study. This task allows the evaluation of specific movements required to produce stop consonants (/p/, /t/, /k/). \cref{tab:dataset} shows the demographic information of each database. 

\begin{table}[]
\caption{Demographic and clinical information of the participants. [F/M]: Female/Male. Values reported as mean $\pm$ std.}
\resizebox{\linewidth}{!}{
\begin{tabular}{lcc}
\hline
\rowcolor[HTML]{C0C0C0} 
                                      & \textbf{PD patients}       & \textbf{HC subjects}        \\ \hline
\rowcolor[HTML]{C0C0C0} 
\multicolumn{3}{c}{\cellcolor[HTML]{C0C0C0}\textbf{Spanish}}                                     \\ \hline
Gender {[}F/M{]}                      & 25/25                      & 25/25                       \\
Age {[}F/M{]}                         & 60.7$\pm$7/61.3$\pm$11     & 61.4$\pm$7/60.5$\pm$12      \\
Range of age {[}F/M{]}                & 49-75/33-81                & 49-76/31-86                 \\
MDS-UPDRS-III {[}F/M{]}               & 37.6$\pm$14/37.8$\pm$22    &                             \\
Speech item (MDS-UPDRS-III) {[}F/M{]} & 1.3$\pm$0.8/1.4$\pm$0.9    &                             \\ \hline
\rowcolor[HTML]{C0C0C0} 
\multicolumn{3}{c}{\cellcolor[HTML]{C0C0C0}\textbf{German}}                                      \\ \hline
Gender {[}F/M{]}                      & 41/47                      & 44/44                       \\
Age {[}F/M{]}                         & 66.2 $\pm$9.7/66.7$\pm$8.7 & 62.6$\pm$15.2/63.8$\pm$12.7 \\
Range of age {[}F/M{]}                & 42-84/44-82                & 28-85/26-83                 \\
UPDRS-III {[}F/M{]}                   & 23.3$\pm$12/22.1$\pm$10    &                             \\
Speech item (MDS-UPDRS-III) {[}F/M{]} & 1.2$\pm$0.5/1.4$\pm$0.6    &                             \\ \hline
\rowcolor[HTML]{C0C0C0} 
\multicolumn{3}{c}{\cellcolor[HTML]{C0C0C0}\textbf{Czech}}                                       \\ \hline
Gender {[}F/M{]}                      & 20/30                      & 20/30                       \\
Age {[}F/M{]}                         & 60.1$\pm$9/65.3$\pm$10     & 63.5$\pm$11/60.3$\pm$12     \\
Range of age {[}F/M{]}                & 41-72/43-82                & 40-79/41-77                 \\
UPDRS-III {[}F/M{]}                   & 18.1$\pm$10/21.4$\pm$12    &                             \\
Speech item (MDS-UPDRS-III) {[}F/M{]} & 0.7$\pm$0.6/0.9$\pm$0.5    &                             \\ \hline
\end{tabular}
}
\label{tab:dataset}
\end{table}

\subsubsection{Feature Extraction} 

To create a representation for each recording, we used Wav2vec 2.0 architecture, a state-of-the-art topology based on transformers proposed in~\cite{wav2vac2}. Wav2Vec 2.0 was trained using a self-supervised pre-training approach that allows the model to learn representations directly from the raw audio signal without additional annotations or labels. The training process involved two main steps. Firstly, the contrastive pre-training, where the model was trained to distinguish between two versions of the same audio signal including a positive sample (a randomly selected segment of the original audio signal) and a negative sample (a randomly selected segment of a different audio signal). The second stage was fine-tuned based on a specific automatic speech recognition (ASR) task. 
Particularly in this work, we used a Wav2Vec 2.0 model, pre-trained on 960 hours of unlabeled audio from the LibriSpeech dataset \cite{7178964}, which was derived from English audiobooks and fine-tuned for ASR on the same audio with the corresponding transcripts. Due to the dynamic representation of 768 dimensions for each array with respect to time, we calculated a static vector for each participant from 6 different statistics (mean, standard deviation (std.), skewness, kurtosis, minimum, and maximum), building a speech representation of 4608 dimensions per recording.

\subsubsection{Federated Learning} 

In order to speed up the collaborative training convergence, the FL process was performed merely for the classification network, i.e., after all the embeddings were locally extracted using the Wav2Vec 2.0 model. Of note, all the data pre-processing and feature extraction steps happened locally by every participating institution without sharing any data with other institutions.

Each institution performed a local training round of the classification network and transmitted the network parameters, i.e., the weights and biases, to a trusted server, which aggregated all the local parameters leading to a set of global parameters. In our implementation, we chose each round to be equal to one epoch of training with the full local dataset.
Afterward, the server transmitted back a copy of the global network to each institution for another round of local training. The process continued until the convergence of the global network. It is worth mentioning that not only each institution did not have access to any training data from others, but also not even to the network parameters of others, rather only an aggregated network, without the knowledge about the contributions of other participating institutions to the global network.
Once the training of the global classification network was converged, every institution could take a copy of the global network and locally utilize it for diagnosing its test data.

\subsubsection{Classification and Evaluation} 

The classification network architecture contained 4 fully-connected layers with different sizes: 1024, 256, 64, and 2, respectively. Rectified linear unit (ReLU) activation and batch normalization \cite{bnmainnpaper} were considered in each layer, and a Softmax activation function was used at the output. 
The fully connected network was trained and evaluated following a stratified 10-fold cross-validation strategy. The process was repeated 5 times 
for a better generalization of the results.
The He initialization scheme \cite{kaimnihepap} was applied to all classification network weights and all the biases were initialized with zeros. 
Cross-entropy was chosen as the loss function and the models were optimized using the Adam optimizer \cite{Kingma2014AdamAM} with a learning rate of $8 \times 10^{-5}$ and weight decays of $5 \times 10^{-6}$.
The classification networks were trained for 50 epochs in batches of size 16.
Accuracy and area under the receiver-operator-characteristic curve (AUC) were chosen as the main evaluation metrics, while sensitivity and specificity were utilized as supporting metrics. 
Two-tailed paired t-test was employed for determining statistical significance. The significance threshold was set at p-value $\le 0.05$.

\section{Experiments and Results}
\label{sec:experiments}

For each test database, we compared the diagnostic performances of the methods in three multicentric setups where the network was: i) locally trained using solely the training set of the corresponding database (Local), ii) trained utilizing the combination of all the training sets of different databases at a central location without privacy measures (Central), and iii) trained with all the training sets of different databases based on FL, i.e., without sharing any data and preserving patient privacy information. 
Furthermore, due to the relatively small test sizes of each database, we repeated each experiment corresponding to each cross-validation step 5 times, including the training and evaluation of the classification network for all 3 setups. Considering the 5 repetitions and 10-fold cross-validation steps, a total of $50$ values were obtained for statistical analysis of each experiment.

\begin{figure*}
    \centering
    \includegraphics[width=\linewidth, trim={0cm, 0cm, 0cm, 0cm}, clip]{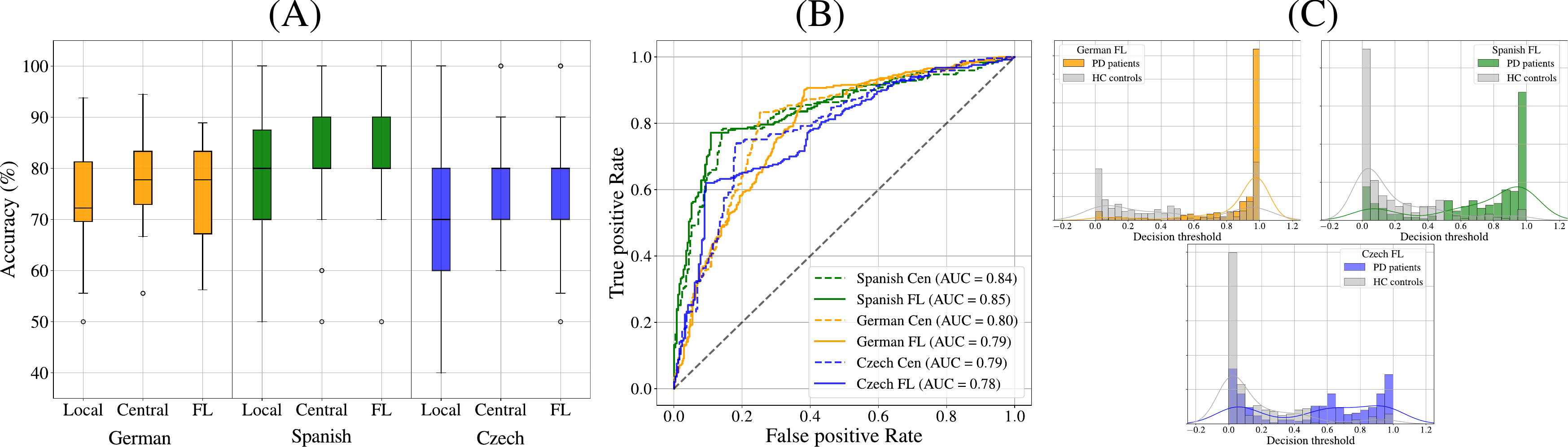}    \caption{Evaluation results. (A) Illustrates the final accuracy values for each test database using the 3 setups, where ``Local" represents solely using the training set of the target database, while ``Central" means utilizing all training sets when combined with each other at a central location. (B) Shows the receiver-operator-characteristic curves. (C) Shows the histogram and the probability density distributions obtained for the classification of German, Spanish, and Czech databases using the FL strategy.}
    \label{fig:results}
\end{figure*}

The average evaluation results are reported in \cref{tab:results} and details about diagnostic accuracy and classification performance are illustrated in \cref{fig:results}-A. The accuracy of the FL method was significantly higher than local models for Spanish ($83.2 \pm 10.8\%$ vs. $77.0 \pm 13.3$; P-value $= 0.001$) and Czech ($76.0 \pm 12.2\%$ vs. $70.3 \pm 14.6$; P-value $=0.020$) databases while it was only slightly higher for the German database ($75.8 \pm 8.3\%$ vs. $74.8 \pm 9.1$; P-value $=0.455$) which contained the largest training set. These results suggest that combining the corpus of the same pathology but in different languages allows generalizing the architecture to classify pathological speech from healthy speech.
Furthermore, comparing the non-private ``Central" and the secure FL strategies, we can observe that the diagnostic accuracy of the FL method was not significantly different from the ``Central" model for Spanish ($83.2 \pm 10.8\%$ vs. $82.0 \pm 11.6$; P-value $= 0.436$) and Czech ($76.0 \pm 12.2\%$ vs. $77.8 \pm 9.2$; P-value $=0.334$) databases while it was for the German database ($75.8 \pm 8.3\%$ vs. $78.9 \pm 8.3$; P-value $=0.023$).
Moreover, \cref{tab:results} shows that the strategy proposed in this work obtained similar results to the state-of-the-art centralized training methods~\cite{vasquez2021transfer}, with the advantage of patient privacy preservation by avoiding data exchange between local institutions using an FL strategy.

\begin{table}[]
\caption{Evaluation results for each database. ``Local" represents solely using the training set of the target database, while ``Central" means utilizing all training sets when combined with each other at a central location. Values are reported as mean $\pm$ std in percentages. The ``P-value" is with respect to FL for each database for accuracy values.}
\resizebox{\linewidth}{!}{
\begin{tabular}{lccccc}
\hline
\textbf{Training set} & \textbf{Accuracy} & \textbf{AUC} & \textbf{Sensitivity} & \textbf{Specificity} & \textbf{P-value} \\
\hline
\rowcolor[HTML]{C0C0C0} 
\multicolumn{6}{c}{\cellcolor[HTML]{C0C0C0}\textbf{Spanish}}                                     \\ \hline
Local & $77.0 \pm 13.3$ & $77.9 \pm 13.1$ & $71.2 \pm 21.4$ & $82.8 \pm 19.4$ & 0.001  \\
Central & $82.0 \pm 11.6$ & $80.6 \pm 12.0$ & $78.4 \pm 17.5$ & $85.6 \pm 17.6$ & 0.436  \\
FL & $83.2 \pm 10.8$ & $83.6 \pm 11.8$ & $77.2 \pm 17.6$ & $89.2 \pm 14.1$ & -  \\ \hline
\rowcolor[HTML]{C0C0C0} 
\multicolumn{6}{c}{\cellcolor[HTML]{C0C0C0}\textbf{German}}                                      \\ \hline
Local & $74.8 \pm 9.1$ & $73.2 \pm 8.6$ & $76.3 \pm 13.8$ & $73.2 \pm 14.2$ & 0.455  \\
Central & $78.9 \pm 8.3$ & $77.5 \pm 8.2$ & $83.1 \pm 13.0$ & $74.8 \pm 17.0$ & 0.023  \\
FL & $75.8 \pm 8.3$ & $77.1 \pm 7.4$ & $90.8 \pm 8.9$ & $60.8 \pm 18.0$ & -  \\ \hline
\rowcolor[HTML]{C0C0C0} 
\multicolumn{6}{c}{\cellcolor[HTML]{C0C0C0}\textbf{Czech}}                                       \\ \hline
Local & $70.3 \pm 14.6$ & $68.4 \pm 14.4$ & $64.8 \pm 26.1$ & $76.4 \pm 23.1$ & 0.020  \\
Central & $77.8 \pm 9.2$ & $77.6 \pm 10.7$ & $74.0 \pm 18.6$ & $82.0 \pm 17.3$ & 0.334  \\
FL & $76.0 \pm 12.2$ & $78.2 \pm 10.7$ & $62.0 \pm 22.6$ & $90.8 \pm 13.5$ & -  \\ \hline
\end{tabular}
}
\label{tab:results}
\end{table}

In addition, \cref{fig:results}-B shows a visual comparison between the ``Central" and the FL strategies from the receiver-operator-characteristic (ROC) curves and the corresponding AUC values obtained in each experiment. Again, when we compared each institution (language) separately, we can observe that the Central and FL curves have the same trend and show no significant differences. It can also be observed that the Spanish language obtains the best result (AUC of 0.85), followed by German (AUC of 0.79) and Czech language (AUC of 0.78). 
Finally, \cref{fig:results}-C shows the histogram and the probability density distributions obtained for the classification of German, Spanish, and Czech databases using the FL strategy.
It can be observed that all three figures have the highest bins at their extremes, which corresponds to a high probability of the decision taken by the classifier. Moreover, it is possible to observe that in the case of Spanish and Czech, the highest bin is for the HC controls, which is related to the reported specificity (89.2\% and 90.8\%, respectively); while for Spanish, the highest bin corresponds to PD patients due to a higher sensitivity (90.8\%).

\section{Discussion}
\label{sec:conclusion}

In this study, we showed the first successful application of cross-language federated learning for PD detection using three pathological speech corpora, including a total of 188 PD and 188 HC subjects, covering Spanish, German, and Czech languages. We used a state-of-the-art topology namely the Wav2vec 2.0 \cite{wav2vac1, wav2vac2} for obtaining speech representations. We compared the performances in three multicentric setups where the architecture was: i) trained locally and separated by language, i.e., monolingual models, ii) trained utilizing the combination of all the training sets at a central location without privacy measures, i.e., cross-lingual model, and iii) trained with all the training sets of different databases based on FL strategy, i.e., without sharing any data and preserving patient privacy.

The results indicated that the FL model outperformed all the local models (mono-lingual models) for every test database in terms of diagnostic accuracy, while not requiring any data sharing between institutions. This result is very interesting and encourages the scientific community to further explore techniques for the generalization of models from databases of the same pathology, in different languages, without the need for sharing information between other institutions (cross-lingual model), which has been a major challenge. In addition, comparing the ``Central" combination and FL strategies, we observed that in the majority of scenarios, the FL method was not significantly different from the Central method in terms of the model's diagnostic accuracy. This shows that the FL paradigm can considerably help the collaboration of institutions around the world in the creation of DL models with large amounts of data, cross-lingual, and preserving patient privacy by avoiding data exchange between local institutions, a major limitation in real-world practice that was not considered in current state-of-the-art cross-lingual approaches.


Our study has limitations. The collaborative FL training process was implemented in a proof-of-concept mode, i.e., using a single institutional network. Due to strict data protection regulations, the implementation of FL among different institutions would be challenging. However, we simulated a realistic setup where every database corresponded to a separate computing entity and we kept the data strictly independent from each other.
As already mentioned, the parameter aggregation mechanism of the central server which was utilized in this study was direct averaging the individual network parameters of each participating database, i.e., the FedAvg algorithm \cite{mcmahan2017communication}, which is the simplest yet the most common aggregation mechanism. Furthermore, the databases utilized in this study were non-independent-and-identically-distributed (non-IID). This was shown to be decreasing the performance of the global model in many different FL applications \cite{nilsson2018performance}.
Consequently, future work could consider more advanced and task-specific aggregation methods such as \cite{li2020federated, deng2020distributionally, feng2022semi} by accounting for the individual contribution of each participating site by analyzing their gradient updates in each FL training round before aggregation that could potentially increase the performance of the global model.
In addition, we considered the most common task of PD detection, i.e., utilizing speech data containing the rapid repetition of the syllables /pa-ta-ka/ for the applicability of FL in pathological speech analysis in this study. In the future, we will extend this by considering further tasks and cross-pathology scenarios.
As a side note, we could conclude that the characterization performed by the Wav2Vec 2.0 method is suitable to model different impairments for PD detection. 
This could be further investigated in the future with other controlled experiments such as at the level of phonemes, words, and phrases that could help interpret the features obtained by this model.

\section{Conclusions}
This paper shows that FL model yields similar or even better results compared to local approaches where mono-lingual models are created for every test database. FL offers the advantage of not requiring any data sharing between institutions, which we hope will encourage researchers and practitioners to improve scientific collaborations among different institutions around the world. The approach shows that FL allows for obtaining competitive results while preserving data privacy.  
We expect these results to promote simpler and more frequent collaborations between medical institutions, and subsequently, to further improve patient outcomes.

\section{Acknowledgments}
STA was supported by the RACOON network under BMBF grant number 01KX2021. JROA and CDRU were funded by UdeA grant number ES92210001. JR was supported by the National Institute for Neurological Research (Programme EXCELES, ID Project No. LX22NPO5107) - funded by the European Union – Next Generation EU. The funders played no role in the design or execution of the study.


\bibliographystyle{IEEEtran}
\bibliography{mybib}

\end{document}